\documentstyle[aps]{revtex}

\begin{document}
\draft
\date{Submitted to Physical Review B, 15 February 1996}

\title {Auger decay of degenerate and Bose-condensed excitons in Cu$_2$O}
\author{G. M. Kavoulakis and Gordon Baym}
\address {Department of Physics, University of Illinois at
Urbana-Champaign, 1110 West Green Steet, Urbana, Illinois 61801}
\maketitle

\begin{abstract}
\baselineskip=14pt

    We study the non-radiative Auger decay of excitons in Cu$_2$O, in which
two excitons scatter to an excited electron and hole.  The exciton decay rate
for the direct and the phonon-assisted processes is calculated from first
principles; incorporating the band structure of the material leads to a
relatively shorter lifetime of the triplet state ortho excitons. We compare
our results with the Auger decay rate extracted from data on highly degenerate
triplet excitons and Bose-condensed singlet excitons in Cu$_2$O.

\end{abstract}

\baselineskip=14pt
\section{Introduction}

    Recent experiments on excitons in Cu$_2$O have been carried out at
sufficiently high densities and low enough temperatures that the Bose-Einstein
statistics they obey becomes significant \cite{snoke}, and, indeed, in
stressed samples, Bose-Einstein condensation of excitons has been observed.  As
we have shown in Ref.  \cite{kbw}, the crucial barrier to condensing excitons
is heating of the exciton gas by the non-radiative {\it Auger decay} process,
in which two excitons collide to produce an ionized electron and hole.  See
Fig.~1.  This process not only heats the gas, but it leads to loss of excitons
as well. [The Auger process is familiar in electron-hole plasmas, in which the
recombination of an electron and a hole excites either an electron high in the
conduction band or a hole deep in the valence band.  This process has been
studied there both theoretically \cite{augplthe} and experimentally
\cite{augplexp}.] In this paper we present detailed calculations of the Auger
process in Cu$_2$O that we used in Ref.  \cite{kbw} to calculate the time
evolution of the exciton gas in Cu$_2$O following laser excitation.

     In the Cu$_2$O experiments, intense pulses of laser
light excite the crystal, creating a gas of (triplet) ortho excitons and
(singlet) para excitons, split by an exchange energy $\Delta E \approx 12$
meV \cite{exch}.  The kinetic energy distribution of ortho excitons as a
function of time from the onset of the laser pulse is observed by spectroscopy
of their photoluminescence (considerably more intense than that of the
para excitons).  In the classical (low density) regime, the energy distribution
is observed to be Maxwell-Boltzmann, described by an effective exciton
temperature.  In the quantum (high density) regime, the spectra
are well fit in terms of an ideal Bose-Einstein
gas with an instantaneous chemical potential, $\mu$, and temperature, $T$.
From these parameters, one can directly calculate the density of the gas.
The experiments of Snoke {\it et al.} observe a ``quantum saturation'' of the
ortho excitons, i.e., a tendency for them to move closely
parallel to the critical line without condensing \cite{snoke}.  The critical
line, an adiabat (constant entropy per particle, $s=S/N$), has the form, $T_c =
(2\pi\hbar^2/mk_B)[n/g\zeta(3/2)]^{2/3}$, where the degeneracy $g=3$ for
ortho excitons and 1 for para excitons.  In general lines in the phase diagram
parallel (in a log-log plot) to the condensation line at higher temperature
are adiabats.

    The important role of the Auger process has been revealed in several
experiments.  The inverse of the total decay time of ortho excitons
in Cu$_2$O shows an approximate proportionality to their density; the
data for non-stressed crystals \cite{augo,augn} is shown in Fig.~2.  In this
figure the quotes indicate that the density plotted on the horizontal axis is
assumed to be proportional to the total radiative recombination luminescence
intensity and is calibrated by comparison to the density deduced from spectral
fits.  The total decay rate is a sum of interconversion of ortho excitons into
para excitons \cite{op} and the Auger process;  the rates of
both processes are comparable.
An additional feature of the Auger process is that it leads to a much shorter
lifetime of ortho excitons; in lightly stressed crystals the
observed ortho-exciton decay time is $\sim$ 0.1 ns \cite{cond} compared
with the para-exciton lifetime at high para-exciton density, $\sim 10^{19}$
cm$^{-3}$, which exceeds 100 ns.
The Auger process also leads to the presence
of ortho excitons in quantum wells, caused by high stress, for times much
longer than the ortho-exciton lifetime; although the exciton gas in the wells
consists primarily of para excitons, high energy electrons and
holes produced in Auger ionization of para excitons lead to reformation of
excitons in essentially random internal angular momentum states \cite{stress}.

    Here we calculate the Auger decay rate for both the direct and
phonon-assisted mechanisms, Figs.~1 and 3. One of our basic conclusions is
that the rate of the phonon-assisted Auger mechanism is much larger than the
rate of the direct process, a consequence of the conduction and the valence
bands having the same parity.  Although the phonon-assisted mechanism requires
participation of a phonon, the rate is enhanced because the process is dipole
allowed.  In Sec.  II, we first derive the matrix element of the direct Auger
process, neglecting for simplicity at this stage the angular momentum
structure of the excitons, then we discuss detailed approximations to the
matrix element and calculate the decay rate for the direct process.  In Sec.
III we generalize the results to take into account the full angular momentum
structure of the ortho and para excitons in Cu$_2$O and explain how specific
properties of the excitons in this material are related to the direct Auger
process.  We then turn, in Sec.  IV, to consider the phonon-assisted Auger
process and discuss its connection with the phonon-assisted radiative
recombination of excitons in Cu$_2$O.  In next section, V, we take into
account effects of the Pauli principle on the decay rate.  Finally, in Sec.
VI, we compare our results with the measured rate and summarize our
conclusions.

\section{Direct process -- calculation of the matrix element and the decay
rate}

    The direct Auger process is described by the sum of the topologically
distinct Feynman diagrams shown in Fig.~1.  In this figure, time progresses
from left to right; the solid lines with forward-going arrows denote
electrons, and those with backward-going arrows denote holes.  The initial
state contains two excitons, of center-of-mass momenta ${\bf K}$ and ${\bf P}$,
described by the wave function
\begin{eqnarray}
   \Psi_i = \Psi_{{\bf K}}({\bf x_e}, {\bf x_h})
   	\Psi_{{\bf P}}({\bf x_e '}, {\bf x_h '}),
\label{24}
\end{eqnarray}
where $\Psi_{{\bf Q}}$ denotes the full wave function of an
exciton of total momentum ${\bf Q}$, and the ${\bf x_i}$ denote electron and
hole coordinates.  We include effects of symmetrization below.  The final
state contains an ionized electron of momentum ${\bf k_e}$ and an ionized hole
of momentum ${\bf k_h}$, described by the wave function
\begin{eqnarray}
   \Psi_f=\Phi_{c,{\bf k_e}}({\bf x_e}) \Phi_{v,{\bf k_h}}({\bf x_h}),
\label{25}
\end{eqnarray}
where the $\Phi_{j,{\bf k}}({\bf x})$ are the normalized Bloch wave functions
for the ionized electron and hole in the final state, and the subscripts $c,v$
denote the conduction and valence bands.  The
wiggly line in the figure denotes the screened Coulomb interaction.  In the
first class of diagrams, A and A$^\prime$, the electron and hole in the same
exciton recombine; in the second class, B and B$^\prime$, particles of
different excitons recombine.

    The exciton wave functions $\Psi_{{\bf Q}}({\bf {x_e}},{\bf{x_h}})$ can be
expressed in terms of the electron and hole Bloch wave functions as
\begin{eqnarray}
         \Psi_{{\bf Q}} ({\bf x_e}, {\bf x_h})= \sum_{{\bf q}}
	\phi_{{\bf q}} \, \Phi_{c, {\bf q} +
	{{\bf Q}}/2} ({\bf x_e}) \Phi_{v, - {\bf q} + {\bf Q}/2} ({\bf x_h}),
\label{22}
\end{eqnarray}
where the Bloch wave functions $\Phi_{j, {\bf k}}({\bf x})$ in band $j$ are
\begin{eqnarray}
  	\Phi_{j,{\bf k}}({\bf x}) = u_{j,{\bf k}}({\bf x})
       e^{i {\bf k} \cdot {\bf x}}.
\label{23}
\end{eqnarray}
The momentum ${\bf q}$ in the $\Phi_{i,{\bf q}}$ in Eq. (\ref{22}) is
restricted to be inside the first Brillouin zone. In general the sum on the
right of this equation should include all the conduction and valence bands,
but for simplicity we neglect all other bands except the lowest conduction
and the highest valence, which is a rather good approximation for the case
of Cu$_2$O.
In the effective mass approximation for the exciton wave function, with
equal electron and hole effective masses (which is approximately true for the
bands we consider), $\Psi_{{\bf Q}}$ takes the form
\begin{eqnarray}
        \Psi_{{\bf Q}} ({\bf{x_e}},{\bf {x_h}}) = \frac 1 {\sqrt \Omega}
            e^{i {\bf Q} \cdot ({\bf x_e} + {\bf x_h}) / 2} \,
            \phi_{\text{rel}}({\bf{x_e}} - {\bf {x_h}}),
\label{21}
\end{eqnarray}
where $\Omega$ is the crystal volume, and $\phi_{\text{rel}}$ is the relative
electron-hole wave function.  Comparing Eqs. (\ref{22}) and (\ref{21}) we see
that $\phi_{{\bf q}}$ is the Fourier transform of $\phi_{\text{rel}}$ times
$\sqrt {\Omega}$; to a good approximation, we may make this identification
even when using complete Bloch waves, Eq. (\ref{23}).

    The interaction between the excitons is, in general, the
dynamically-screened Coulomb interaction, $V_{\text{eff}}({\bf k},
\omega)=V_{\text{Coulomb}}({\bf k})/ \epsilon({\bf k},\omega)$,
where the dielectric function is given by the usual Lindhard result
\begin{eqnarray}
  \epsilon({\bf k},\omega) = \epsilon_{\infty} - V({\bf k}) \sum_{{\bf q},l,l'}
   \frac {n_{l',{\bf k} + {\bf q}} - n_{l,{\bf q}}}
  {\varepsilon_{l',{\bf k} + {\bf q}} - \varepsilon_{l,{\bf q}} - \hbar \omega}
   | \langle u_{l',{\bf k} + {\bf q}} \mid u_{l, {\bf q}} \rangle |^2 ,
\label{AI.1}
\end{eqnarray}
with $\epsilon_{\infty}$ the high frequency dielectric constant of the
material, and $l$ and $l'$ band indices.  The second term in
(\ref{AI.1}) can in fact be neglected in the Auger process, where we are
interested primarily in small momentum transfers of order the thermal momentum
of excitons, $P^2/2M_{\text{exc}}$ on the scale of 1-10 meV, but rather large
frequencies corresponding to energy transfer on the order of the energy gap,
$\hbar \omega \approx E_g=2.17$ eV.  In this limit, the terms in Eq.
(\ref{AI.1}) involving the same band vanish for finite $\omega$. In the same
limit, for different bands, $l \neq l'$, the last term on the right side of Eq.
(\ref{AI.1}) gives a $k$-independent contribution which can be ignored for
low densities. Thus $\epsilon({\bf k},\omega_g) \approx \epsilon_{\infty}$,
and the effective interaction, essentially independent of
$\omega$ for large $\omega$, becomes
\begin{eqnarray}
  V_{\text{eff}}({\bf k}) = \frac {4\pi e^2}{\Omega\epsilon_{\infty}k^2}.
\label{AI.3}
\end{eqnarray}
We note that as a consequence of Cu$_2$O being almost non-polar
\cite{okeefe}, $\epsilon_{\infty} \, (\approx 6.46)$ and the static dielectric
constant $\epsilon_0 \, (\approx 7.11)$ differ only slightly.

    Assembling these results, we find that the matrix element
$M \equiv \langle \Psi_{f} | V|\Psi_{i} \rangle$
    between the
initial and final states, for diagram A for example, is thus
\begin{eqnarray}
  M_A   = \sum_{{\bf q},{\bf G},{\bf G'}} \phi_{-{\bf k_h}+{\bf P}/2}
           V_{\text{eff}}({\bf G} - {\bf K}) \, \phi_{{\bf q}}
  \langle u_{v,{\bf q} - {\bf K}/2} | u_{c, {\bf q} + {\bf K}/2}
  \rangle_{{\bf G}}
    \langle u_{c,{\bf k_e}} | u_{c, {\bf k_e} + {\bf K}} \rangle_{{\bf G'}} \,
\delta_{{\bf k_e} + {\bf k_h} +{\bf G} + {\bf G'}, {\bf K} + {\bf P}},
\label{26}
\end{eqnarray}
where
\begin{eqnarray}
   \langle u_{v,{\bf q} - {\bf K}/2} | u_{c, {\bf q} + {\bf K}/2}
   \rangle_{{\bf G}}
   \equiv \int d {\bf x} \, u_{v,{\bf q} - {\bf K}/2}^{*} ({\bf x})
  u_{c,{\bf q} + {\bf K}/2} ({\bf x}) e^{i {\bf G} \cdot {\bf x}}
\label{28}
\end{eqnarray}
is a generalized overlap integral \cite{Land} between the conduction and the
valence bands. The momenta ${\bf p}$ in $\phi_{{\bf p}}$ are restricted
to the first Brillouin zone.  The ${\bf x}$-integral is over a unit cell,
and the summation is over all reciprocal lattice vectors ${\bf G},{\bf G'}$.
Energy and crystal momentum conservation restrict this summation considerably.
Since $K$ and $P$ are determined by the thermal motion of excitons they are
negligible compared to $k_e$ and $k_h$, which are determined by the scale of
the energy gap $E_g$ and the reciprocal lattice vectors.  In this limit the
condition for conservation of crystal momentum in Eq.  (\ref{26}) becomes,
\begin{eqnarray}
  {\bf k_e} + {\bf k_h} + {\bf G} + {\bf G'} \approx 0.
\label{ap1}
\end{eqnarray}
In the same limit the energy conservation condition,
\begin{eqnarray}
 \frac {\hbar^2 k_e^2} {2 m_e} + \frac {\hbar^2 k_h^2} {2 m_h} = E_g,
\label{ap2}
\end{eqnarray}
puts an upper bound on the quanity $|{\bf k_e} + {\bf k_h}|$.  Numerically,
$|{\bf k_e} + {\bf k_h}| a_{\ell} \le 5.5$, where $a_{\ell}$ is the
lattice constant.  Since for the smallest nonzero reciprocal lattice vector
${\bf G_0}$, $G_0 a_{\ell} = 6.28$, the conservation laws are
satisfied only for
\begin{eqnarray}
 {\bf G} + {\bf G'} = 0.
\label{ap3}
\end{eqnarray}

    Using the fact that for a relative 1s electron-hole state, $\phi_{-{\bf q}}
= \phi_{{\bf q}}$, the final expression for the sum of the
matrix elements for processes A and A$^\prime$ in Fig.~1 is,
\begin{eqnarray}
     M_A+M_{A'} &=& \sum_{{\bf q},{\bf G}}
                        V_{\text{eff}}({\bf G} - {\bf K})
                    \phi_{{\bf q}} \langle u_{v,{\bf q} - {\bf K}/2} |
                u_{c,{\bf q} + {\bf K}/2} \rangle_{{\bf G}}
   \,   \delta_{{\bf k_e} + {\bf k_h}, {\bf K} + {\bf P}}
\phantom{XXXXXXXXXXXXXXXXXXXX}
   \nonumber\\
    &\times& \left( \phi_{{\bf k_h}-{\bf P}/2}
   \langle u_{c,{\bf k_e}} | u_{c, {\bf k_e} + {\bf K}} \rangle_{-{\bf G}}
   - \phi_{{\bf k_e}-{\bf P}/2}
   \langle u_{v,{\bf k_h}} | u_{v, {\bf k_h} + {\bf P}} \rangle_{-{\bf G}}
\right).
\label{36}
\end{eqnarray}
The first term in parentheses in Eq.  (\ref{36}) corresponds
to process A and the second to process A$^\prime$, which enters with a change
of sign because the Coulomb exchange couples to the positively charged hole
instead of the electron.

    In the second class of diagrams, processes B and B$^\prime$
are equivalent to processes A and A$^\prime$ with either the two electrons or
the two holes interchanged.  For this reason these processes enter with a
minus sign relative to A and A$^\prime$, and we find
 \begin{eqnarray}
     M_B + M_{B'} =
      - \sum_{{\bf q},{\bf G}}
     \phi_{\bf q} \,
     \delta_{{\bf k_e} + {\bf k_h},{\bf K} + {\bf P}}
    \phantom{XXXXXXXXXXXXXXXXXXXXXXXXXXXXXXX}
    \nonumber\\
     \times  \left( \phi_{{\bf k_h}-{\bf P}/2}
    V_{\text{eff}}({\bf k_e}+{\bf G}-{\bf q}-{\bf K}/2)
   \langle u_{v,{\bf q}-{\bf K}/2} | u_{c,{\bf k_e}-{\bf K}} \rangle_{-{\bf G}}
  \langle u_{c,{\bf k_e}} | u_{c, {\bf q} + {\bf K}/2} \rangle_{{\bf G}}
\right.
  \phantom{XXXX}
    \nonumber \\
   \left. - \phi_{{\bf k_e} - {\bf K}/2}
   V_{\text{eff}}({\bf k_h} + {\bf G} - {\bf q} - {\bf P}/2)
  \langle u_{v,{\bf k_h} - {\bf P}} | u_{c,{\bf q} - {\bf P}/2}
  \rangle_{-{\bf G}}
 \langle u_{v,{\bf k_h}} | u_{v, {\bf q} + {\bf P}/2} \rangle_{\bf G}
\right) .
 \phantom{XXXX}
\label{38}
\end{eqnarray}

    We now consider the effects of the band structure of Cu$_2$O on the
generalized overlap integral, Eq.  (\ref{28}).  For ${\bf G} = 0$,
(\ref{28}) is the usual overlap integral between the conduction and the
valence bands \cite{overl}.  We note that for ${\bf K} = 0$, the wave vector of
condensed excitons, the overlap integral for ${\bf G} = 0$ vanishes as a
consequence of orthogonality of Bloch wave functions between different bands.

    The excitons in Cu$_2$O in the condensation experiments under study are
formed from electrons in the lowest conduction band ($\Gamma_6^{+}$) and holes
in the highest valence band ($\Gamma_7^{+}$) (the ``yellow series'').  The
fact that these two bands have even parity \cite{bstr} implies that
\begin{eqnarray}
    u_{j,{\bf k}} (-{\bf x}) = u_{j,-{\bf k}} ({\bf x}).
\label{34}
\end{eqnarray}
This result is readily seen in the tight-binding approximation for the
Bloch wave functions for which
\begin{eqnarray}
 	\Phi_{j,{\bf k}}({\bf x})=\sum_{{\bf R}} \Phi_{a,j}({\bf x} -{\bf R})
 \, e^{i {\bf k} \cdot {\bf R}},
\label{32}
\end{eqnarray}
where the summation is over all crystal points and the $\Phi_{a,j}({\bf x})$
are the atomic orbitals which form the band with index $j$; for
positive parity,
\begin{eqnarray}
    \Phi_{j,{\bf k}} (-{\bf x}) = \Phi_{j, -{\bf k}}({\bf x}),
\label{33}
\end{eqnarray}
from which (\ref{34}) follows. Thus, for the ${\bf G} = 0$ term
in the sum over ${\bf q}$ in the matrix elements $M_A$ and $M_{A'}$,
\begin{eqnarray}
      \sum_{{\bf q}} \phi_{{\bf q}}
	\langle u_{v,{\bf q} - {\bf K}/2} | u_{c, {\bf q} + {\bf K}/2}
      \rangle_{{\bf G}=0} =
	\sum_{{\bf q}} \phi_{{\bf q}}
  \langle u_{v,{\bf q} + {\bf K}/2} | u_{c, {\bf q} -{\bf K}/2}
  \rangle_{{\bf G}=0},
\label{35}
\end{eqnarray}
which implies that the sum is an even function of $K$
and is thus proportional to $K^2$ (since it vanishes for
$K=0$). We use ${\bf k} \cdot {\bf p}$ perturbation theory \cite{kpa} to
calculate the overlap integral in the above matrix element. For example,
\begin{eqnarray}
        |u_{c, {\bf q} + {\bf K}/2} \rangle
    \approx |u_{c, {\bf q} } \rangle +
   \frac {\hbar} m \sum_{i \ne c} \frac {\langle u_{i, {\bf q} }|
  {\bf K}/2 \cdot {\bf p} \,|
 u_{c,{\bf q} } \rangle } {\varepsilon_{c, {\bf q} } -
\varepsilon_{i, {\bf q} }}
|u_{i, {\bf q} } \rangle,
\label{62}
\end{eqnarray}
where $m$ is the bare electron mass.  The sum is over all the negative
parity bands of Cu$_2$O.  Of the ten valence and four conduction bands in this
material, only the conduction band $(c')$ that is $\approx$ 449 meV higher than
the $\Gamma_6^{+}$ conduction band $(c)$ and the one very deep valence band
$(v')$ $\approx$ 5.6 eV below the $\Gamma_7^{+}$ valence band $(v)$ have odd
parity, Fig. 6. The contribution of the odd-parity conduction band is expected
to dominate the sum because of the small energy denominator in Eq.  (\ref{62}).
Thus
\begin{eqnarray}
      \langle u_{v,{\bf q} - {\bf K}/2}|
     u_{c, {\bf q} + {\bf K}/2 }\rangle \approx
    \left( \frac \hbar m \right)^2 \sum_{i \neq c,v} \frac
   {\langle u_{v, {\bf q} }| - {\bf K}/2 \cdot {\bf p}  |
  u_{i, {\bf q} } \rangle  \langle u_{i, {\bf q} }| {\bf K}/2 \cdot {\bf p}
  |  u_{c, {\bf q} } \rangle } {(\varepsilon_{c, {\bf q} } -
 \varepsilon_{i, {\bf q} })
(\varepsilon_{i, {\bf q} } -\varepsilon_{v, {\bf q} })}.
\label{63a}
\end{eqnarray}

   The ${\bf G} = 0$ term (which we denote by ``normal'') in the sum (\ref{36})
for the matrix elements
$M_A + M_{A'}$, in the limit $K,P \ll k_e,k_h$, is equal to
\begin{eqnarray}
  (M_A + M_{A'})_{\text{normal}} &=& \frac {4 \pi e^2}
  {\Omega \epsilon_{\infty} K^2} (\phi_{{\bf k_e} -{\bf K} -{\bf P}/2} -
  \phi_{{\bf k_e} -{\bf P}/2})  \sum_{{\bf q}}  \phi_{{\bf q}} \,
   \langle u_{v,{\bf q} - {\bf K}/2} | u_{c,{\bf q} + {\bf K}/2} \rangle \,
 \delta_{{\bf k_e} + {\bf k_h}, {\bf K} + {\bf P}}     \\ \nonumber
&\approx& \frac {4 \pi e^2}  {\Omega \epsilon_{\infty} K^2} \,
  {\bf K} \cdot
   \left( \frac {\partial \phi_{{\bf k}}}{\partial {\bf k}} \right)_{{\bf k_e}}
  \sum_{{\bf q}} \phi_{{\bf q}} \,
    \langle u_{v,{\bf q} - {\bf K}/2} | u_{c,{\bf q} + {\bf K}/2} \rangle \,
   \delta_{{\bf k_e} + {\bf k_h}, {\bf K} + {\bf P}}.
\label{37e}
\end{eqnarray}
Taking $\phi_{\text{rel}}(x)$ to be the hydrogenic ground state wave function
with excitonic Bohr radius $a_B$ ($\approx 7$ \AA) and using Eq.
(\ref{63a}), we estimate the above quantity as
\begin{eqnarray}
  (M_A + M_{A'})_{\text{normal}} \approx
           - \frac {2^7 \pi e^2} {\Omega \epsilon_{\infty}}
             \frac {({\bf k_e} a_B) \cdot ({\bf K} a_B)} {(1+(k_e a_B)^2)^3}
             \frac {\hbar^2} {m^2} \sum_{i \neq c,v}
        \frac {({\bf K} \cdot {\bf p}_{v,i}) ({\bf K} \cdot {\bf p}_{i,c})}
        {K^2 (\varepsilon_{c,0} - \varepsilon_{i,0})
             (\varepsilon_{i,0} - \varepsilon_{v,0})}
        \, \delta_{{\bf k_e} + {\bf k_h}, {\bf K} + {\bf P}},
\label{38e}
\end{eqnarray}
where ${\bf p}_{i,j}$ is the matrix element of the momentum operator
between the Bloch states of the bands $i$ and $j$ at the zone center.  The
matrix elements ${\bf p}_{i,j}$ can be extracted from experiment;
if we assume, for example,
that they are all of equal magnitude, then Ref.  \cite{Niki} implies that
$|{\bf p}_{i,j}|/\hbar \approx 0.13$ \AA$^{-1}$.

    If $K$ and $P$ are ignored compared to $k_e$ and $k_h$, in the case that
${\bf G}=0$, the right side of Eq. (\ref{36}) vanishes. The contribution of
Umklapp processes to Eq. (\ref{36}) is,
\begin{eqnarray}
    (M_A+M_{A'})_{\text{Umklapp}} \approx \sum_{{\bf q}, {\bf G} \neq 0}
	V_{\text{eff}}({\bf G}) \, \phi_{{\bf q}} \,
	\phi_{{\bf k_e}} \langle u_{v,{\bf q}} | u_{c,{\bf q}} \rangle_{{\bf G}}
     \left( \langle u_{c,{\bf k_e}} | u_{c,{\bf k_e}} \rangle_{-{\bf G}} -
   \langle u_{v,{\bf k_h}} | u_{v,{\bf k_h}} \rangle_{-{\bf G}} \right)
  \delta_{{\bf k_e} + {\bf k_h}, {\bf K} + {\bf P}},
\label{37}
\end{eqnarray}
which is nonzero, but very small. In the nearly free electron model,
for example, with $u_{v,{\bf k}}({\bf x}) = 1/ \sqrt{\Omega_c}$ and
$u_{c,{\bf k}}({\bf x}) = e^{i {\bf G_0} \cdot {\bf x}} / \sqrt{\Omega_c}$,
where $\Omega_c$ is the volume of the unit cell and ${\bf G_0}$ one of the
six smallest reciprocal lattice vectors, the right side of Eq. (\ref{37})
vanishes.

    We turn now to the processes B and B$^\prime$.  In the limit $K, P \ll
k_e, k_h$ Eq.  (\ref{38}) reduces to
\begin{eqnarray}
     M_B+M_{B'} \approx - \sum_{{\bf q},{\bf G}}
	\phi_{{\bf q}} \, \delta_{{\bf k_e} + {\bf k_h},{\bf K}+{\bf P}} \,
   \left(\phi_{{\bf k_h}} V_{\text{eff}}({\bf k_e} + {\bf G} - {\bf q})
	\langle u_{c, {\bf k_e}} | u_{c,{\bf q}} \rangle_{{\bf G}}
     \langle u_{v, {\bf q}} | u_{c,{\bf k_e}} \rangle_{-{\bf G}}
     \phantom{XX}
\right.
    \nonumber\\
     -\left. \phi_{{\bf k_e}}
   V_{\text{eff}}({\bf k_h} +{\bf G} - {\bf q})
   \langle u_{v,{\bf k_h}} |
  u_{c,{\bf q}} \rangle_{-{\bf G}}
 \langle u_{v, {\bf k_h}} | u_{v,{\bf q}} \rangle_{{\bf G}}
\right).
\label{39}
\end{eqnarray}
For ${\bf G} \neq 0$, in the same nearly free electron approximation,
the right side of Eq. (\ref{39}) vanishes. Since the overlap integrals
are less than unity in magnitude,
an upper bound of the ${\bf G}=0$ term is given by
\begin{eqnarray}
 (M_B + M_{B'})_{\text{normal}} \leq  2 \phi_{{\bf k_e}}
 \sum_{{\bf q}} \phi_{{\bf q}} V_{\text{eff}}({\bf k_e} - {\bf q})
 \langle u_{v,{\bf q}} | u_{c,{\bf k_e}} \rangle \,
     \delta_{{\bf k_e} + {\bf k_h}, {\bf K} + {\bf P}}.
\label{39a}
\end{eqnarray}
The overlap integral above is approximately
\begin{eqnarray}
         \langle u_{v,{\bf q}} | u_{c,{\bf k_e}} \rangle \approx
        \left( \frac {\hbar} m \right)^2 \sum_i
       \frac {({\bf q} - {\bf k_e}) \cdot {\bf p}_{c,i} \,
      ({\bf q} - {\bf k_e}) \cdot {\bf p}_{i,v}}
     {(\varepsilon_{c,0} -\varepsilon_{i,0})
    (\varepsilon_{i,0} - \varepsilon_{v,0})},
\label{39ab}
\end{eqnarray}
where $q$ ranges approximately between $k_e - 1/a_l$ and $k_e + 1/a_l$.
Then,
\begin{eqnarray}
 (M_B + M_{B'})_{\text{normal}} \leq 41
       \frac {e^2} {\Omega \epsilon_{\infty}}
      \frac 1 {[1+(k_e a_B)^2]^2}
     \left( \frac {\hbar} m \right)^2 \sum_i
    \frac {|{\bf p}_{c,i}| |{\bf p}_{i,v}|}
   {(\varepsilon_{c,0}-\varepsilon_{i,0})(\varepsilon_{i,0}-\varepsilon_{v,0})}
\,
  \delta_{{\bf k_e} + {\bf k_h}, {\bf K} + {\bf P}},
\label{39b}
\end{eqnarray}
where the numerical factor is the result of a dimensionless integral times
constants. We show below using Eq. (\ref {39b}) that processes B and
B$^\prime$ contribute negligibly to the Auger decay rate.
Only A and A$^\prime$ contribute significantly to the direct
Auger mechanism; these processes do not allow condensed excitons to Auger
recombine, since $M_A+M_{A'}$ vanishes for $K=0$ (Eq. (\ref{38e})).

    We next proceed to calculate the decay rate $\Gamma_{{\bf K},{\bf P}}$ for
the direct process in which two excitons of momentum ${\bf K}$ and ${\bf P}$
collide. From the golden rule
\begin{eqnarray}
   \Gamma_{{\bf K}, {\bf P}} = \frac {2 \pi} {\hbar}
   \sum_{{\bf k_e}, {\bf k_h}} |M|^2
   (1-n_{c,{\bf k_e}}) (1-n_{v,{\bf k_h}}) \,
   \delta(E_{{\bf K}}+E_{{\bf P}}-
   \varepsilon_{c, {\bf k_e}} - \varepsilon_{v, {\bf k_h}}),
   \label{51}
\end{eqnarray}
where $M$ is the matrix element for the process.  The exciton energies are
\begin{eqnarray}
         E_{{\bf Q}}=E_g-E_b+ \frac {\hbar^2 Q^2} {2 M_{\text{exc}}},
         \label{52a}
\end{eqnarray}
where $M_{\text{exc}}$ is the exciton mass,
$E_b$ is the binding energy, $\approx 141$ meV including the exchange energy
$\Delta E \approx 12$ meV which is nonzero for the ortho excitons only
(see Sec. III). We may neglect $E_b$ and $\Delta E$ compared to the energy
gap $E_g$.  The electron and the hole energies are
\begin{eqnarray}
 \varepsilon_{c, {\bf k_e}}=E_g + \frac {\hbar^2 k_e^2} {2 m_e},
       \phantom{X}
      \varepsilon_{v, {\bf k_h}}= \frac {\hbar^2 k_h^2} {2 m_h},
\label{52}
\end{eqnarray}
and the $n_{j,{\bf k_i}}$'s are the electron and hole occupation numbers
in the final ionized states.  Since these final states have very high
energy, we may neglect the $n_{j,{\bf k_i}}$ in Eq.  (\ref{51}), as well as the
inverse process.

    The decay rate per unit volume for the direct Auger process,
$\Gamma_{A,d}/\Omega$, is given by
\begin{eqnarray}
       \frac{1}{\Omega}\Gamma_{A,d} =
    - \frac{1}{\Omega} \left( \frac {\partial N} {\partial t} \right)_{A,d} =
     \frac{1}{\Omega}
\sum_{{\bf K},{\bf P}} f_{{\bf K}} f_{{\bf P}} \Gamma_{{\bf K},{\bf P}} \equiv
   \frac n {\tau_{A,d}} ,
\label{53}
\end{eqnarray}
where $f_{{\bf K}}$ is the distribution function of excitons,
$n$ the density of excitons and $\tau_{A,d}$ the scattering time for
the direct Auger process.

     Using Eq. (\ref{38e}) we estimate the contribution to the decay rate due
to the (dominant) processes A and A$^\prime$
\begin{eqnarray}
          \tau_{A,d}^{-1} \approx 0.77 \times 2^{17} \pi \,
         \frac {\mu e^4} {2 \hbar^3 \epsilon_{\infty}^2}
  \frac {(k_g a_B)^3} {[(k_g a_B)^2+1]^6} \frac {M_{\text{exc}} a_B^2} {\hbar^2}
        \left( \frac {\hbar} {m a_B} \right)^4
       \sum_i \left( \frac {|{\bf p}_{c,i}| |{\bf p}_{i,v}|}
      {(\varepsilon_{c,0} - \varepsilon_{i,0})
     ( \varepsilon_{i,0} - \varepsilon_{v,0})} \right)^2
    \, k_B T
   \,\, n_{\text{exc}} a_B^3,
\label{55}
\end{eqnarray}
where $E_g \equiv {\hbar}^2 k_g^2 /2 \mu$, $\mu$ is the reduced mass
for the electron and the hole,
$n_{\text{exc}}$ is the density of noncondensed (excited)
excitons and $T$ is the temperature of the exciton gas. We note that
$\tau_{A,d}$ is inversely proportional to the density of noncondensed
excitons. For the proportionality constant for excitons in Cu$_2$O,
\begin{eqnarray}
      \tau_{A,d}^{-1} \approx 5 \times 10^{-4} \,
     n_{\text{exc}} \, T \, \text{ns}^{-1},
\label{55a}
\end{eqnarray}
with $T$ measured in kelvin and $n_{\text{exc}}$ measured in units of $10^{18}
\text{cm}^{-3}$. As we will see this is
decay rate is small compared to the rate for the phonon-assisted process,
but it can contribute to para-exciton para-exciton Auger collisions (which are,
as we will see, forbidden in the zero-stress case) in highly stressed crystals.
We explain this effect in the next section.

 Finally Eq. (\ref{39b}) gives an upper bound for $\tau_{A,d;B+B'}^{-1}$, i.e.,
the contribution of processes B and B$^\prime$ to $\tau_{A,d}^{-1}$,
\begin{eqnarray}
     \tau_{A,d;B+B'}^{-1} \le \frac {2 \times 41^2} {\pi}
  \frac {\mu e^4} {2 \hbar^3 \epsilon_{\infty}^2}
     \frac {(k_g a_B)} {[1+(k_g a_B)^2]^4}
       \left( \frac {\hbar} {m a_B} \right)^4
	\sum_i \left( \frac {|{\bf p}_{c,i}| |{\bf p}_{i,v}|}
    {(\varepsilon_{c,0}-\varepsilon_{i,0})
	 (\varepsilon_{i,0}-\varepsilon_{v,0})} \right)^2
 \, n a_B^3.
\label{54a}
\end{eqnarray}
For the parameters of excitons in Cu$_2$O, Eq. (\ref{54a}) gives
$\tau_{A,d;B+B'}^{-1} \le 4 \times 10^{-4} \, n \, \text{ns}^{-1}$,
with $n$ measured in units of $10^{18} \text{cm}^{-3}$, which is negligible
compared with the contribution of the processes A and A$^\prime$ as well
as the most important processes of the phonon-assisted Auger mechanism.

\section{Spin statistics -- Band structure
                and optical properties of Cu$_2$O}

      In this section we review the band structure underlying the properties
of excitons in Cu$_2$O and their direct Auger recombination process.
These properties infuence the direct as well as the phonon-assisted mechanisms.
We discuss the effects of the band structure on the phonon-assisted Auger
process in Sec. IV.

     The direct radiative recombination of both ortho and para excitons
is dipole forbidden because the conduction and valence bands have the
same parity \cite{bstr,Nik}. The quadrupole recombination of ortho excitons
is allowed and has been observed experimentally \cite{Gross}.
In contrast, the direct radiative recombination of para excitons is highly
forbidden \cite{Mys}, but becomes allowed if uniaxial stress is applied
\cite{pstr}. Both species can recombine via phonon-assisted processes.

     The exchange interaction, shown in Fig. 4, where an
exciton virtually annihilates and reforms, is responsible for the ortho-para
energy splitting at the zone center of Cu$_2$O. Although for pure spin states
this interaction is nonzero for the singlet, shifting the energy
of the singlet state higher than the triplet, experimentally the ortho excitons
lie higher than the para excitons by 12 meV \cite{exch}. The absorption
spectrum of Cu$_2$O \cite{Nik,OPA} shows discrete lines below the
continuum which correspond to excitonic absorption, but the n=1 line
is very weak, in contrast with the n=2,3,... lines.
All these observations can be understood in terms of the band structure
of Cu$_2$O \cite{bstr}. The conduction band is formed by Cu 4s
orbitals and the valence band by Cu 3d orbitals. The fivefold
degenerate (without spin) Cu 3d orbitals split under the crystal field
into a higher threefold $\Gamma_{25}^{+}$ and a lower twofold degenerate band
$\Gamma_{12}^{+}$. Finally, $\Gamma_{25}^{+}$ splits further
because of the spin-orbit interaction into two bands, a higher $\Gamma_{7}^{+}$
non-degenerate band and a lower twofold degenerate $\Gamma_{8}^{+}$
band (Fig. 5). The $\Gamma_{7}^{+}$ and $\Gamma_{8}^{+}$ mix with the
$\Gamma_6^{+}$ conduction band to form the yellow and green exciton series,
respectively.

     From now on we restrict our discussion to the yellow exciton series.
The weakness of the n=1 line in the absorption spectrum is due to the
$\Gamma_6^{+}$ and $\Gamma_{7}^{+}$ bands having the same parity,
making the transition to this line only quadrupole-allowed.
In contrast to the n=1 line, the n=2,3,... lines are dipole-active,
provided that the created excitons are in p-states.
The total angular momentum functions for the exciton triplet states are
\begin{eqnarray}
         &|J=1, J_z=1 \rangle& = |\uparrow_e, \uparrow_H \rangle
\label{41} \\
       &|J=1, J_z=0 \rangle& = \frac 1 {\sqrt 2}
      \left( {|\uparrow_e, \downarrow_H \rangle -
    |\downarrow_e, \uparrow_H \rangle } \right)
\label{42} \\
     &|J=1, J_z=-1 \rangle& = |\downarrow_e, \downarrow_H \rangle,
\label{43}
\end{eqnarray}
and for the singlet states,
\begin{eqnarray}
          |J=0, J_z=0 \rangle = \frac 1 {\sqrt 2}
  \left( {|\uparrow_e, \downarrow_H \rangle +
 |\downarrow_e, \uparrow_H \rangle} \right) .
\label{44}
\end{eqnarray}
The indices $e,H$ refer to the electron and the hole, respectively: while
the electron states are pure spin states, the hole states are {\it total\/}
angular momentum states,
\begin{eqnarray}
    |\uparrow_H \rangle = - \frac 1 {\sqrt 3} \left[ (X+iY)
\,  |\downarrow_h \rangle
 + Z \, |\uparrow_h \rangle \right]
\label{45}
\end{eqnarray}
and
\begin{eqnarray}
     |\downarrow_H \rangle = - \frac 1 {\sqrt 3} \left[ (X-iY)
\, 	|\uparrow_h \rangle
 - Z \, |\downarrow_h \rangle \right] ,
\label{46}
\end{eqnarray}
where the states with lower case $h$
are pure spin states. The spatial functions $X,Y,Z$ transform
as yz, xz and xy, respectively.

     Using these angular momentum functions we explain now the
experimental observations mentioned above,
turning to the direct Auger decay process at the end of the section.
In the radiative recombination of an exciton the electron goes from
the conduction to the valence band with emission of a photon.
If we assume that there is no spin-flip (a higher-order effect),
the above angular momentum functions imply that the matrix element
for the direct radiative recombination rate for the ortho excitons
is proportional to $\sqrt {2/3}$ times the result of the spatial part of the
calculation, but that it vanishes for the para excitons. Similarly,
para excitons are not allowed to recombine directly and the exchange
interaction vanishes for the singlet. For the triplet state the exchange
interaction is nonzero and raises the ortho-exciton energy with respect to the
para-exciton energy; the energy splitting $\Delta E$ at the zone center
($K=0$) is given by \cite{exch}
\begin{eqnarray}
    \Delta E = \frac 2 3 \int \Psi_{{\bf K}=0}({\bf x},{\bf x})
     V({\bf x} - {\bf x'})
      \Psi_{{\bf K}=0}^{*}({\bf x'},{\bf x'}) \, d {\bf x} \, d {\bf x'},
\label{47}
\end{eqnarray}
where $V({\bf x} - {\bf x'})$ is the Coulomb potential; the factor of $2/3$
again comes from the angular momentum wave functions. Using Eq. (\ref{22})
for the exciton wave function we find that
\begin{eqnarray}
   \Delta E = \frac 2 3 \sum_{{\bf q},{\bf p},{\bf G} \neq 0} \frac {4 \pi e^2}
        {\Omega |{\bf G}|^2 \epsilon_{\infty}}
      \phi_{{\bf q}} \langle u_{v,{\bf q}} | u_{c,{\bf q}} \rangle_{{\bf G}}
    \phi_{{\bf p}}^{*}
   \langle u_{v,{\bf p}} | u_{c,{\bf p}} \rangle_{{\bf G}}^*.
\label{48}
\end{eqnarray}
The sum is dominated by the terms with smallest ${\bf G}$'s
(six in number because of the cubic symmetry of the crystal), which we denote
by $G_0$; then
\begin{eqnarray}
 \Delta E \, {\raisebox{-.5ex}{$\stackrel{<}{\sim}$}} \,
 \frac {12} 3 \, \frac {4 \pi e^2} {\Omega \epsilon_{\infty}}
 \Omega |\phi_{\text{rel}}(0)|^2 \frac {a_{\ell}^2} {4 \pi^2}
     |\langle u_{v,{\bf 0}} | u_{c,{\bf 0}} \rangle_{{\bf G_0}}|^2
      = \frac {e^2} {a_B \epsilon_{0}}
    \frac 4 {\pi^2} \frac {\epsilon_0} {\epsilon_{\infty}}
   \left( \frac {a_{\ell}} {a_B} \right)^2
   |\langle u_{v,{\bf 0}} | u_{c,{\bf 0}} \rangle_{{\bf G_0}}|^2.
\label{49}
\end{eqnarray}
Since experimentally $\Delta E \approx 12$ meV \cite{Birm}, we have
$|\langle u_{v,{\bf 0}} | u_{c,{\bf 0}} \rangle_{{\bf G_0}}| \approx 0.62$.

      In Auger decay, processes B and B$^\prime$ are not affected
by the angular momentum of the electron and the hole. The
forbiddenness of the direct radiative recombination of the
para excitons due to the spin conservation implies that the recombination
vertices which appear in processes A and A$^\prime$ of Fig.~1 are nonzero
{\it only\/} for an ortho exciton decaying and ionizing with {\it either\/} an
ortho exciton or a para exciton.

     Uniaxial stress mixes the $\Gamma_7^+$ and $\Gamma_8^+$ valence bands
with the results that the para-exciton energy increases for small values of
the applied stress but decreases for higher values, and the
direct radiative recombination increases quadratically with increasing stress
\cite{pstr}. In the Auger process this mixing implies that
the recombination vertex of stressed para excitons becomes allowed.
Para excitons in highly-stressed crystals are therefore able to Auger recombine
(processes A and A$^\prime$), ionizing either an ortho exciton or a
para exciton, just like the ortho excitons in the unstressed case.
We discuss this effect further in Sec. VI.

\section{Phonon-assisted Auger process}

    The phonon-assisted Auger mechanism with the participation of a
longitudinal, odd-parity optical phonon is in fact the dominant Auger process.
Although this process requires a phonon and its rate is therefore reduced by
factors of the exciton-phonon interaction, the dipole matrix element between
the intermediate state and the conduction or the valence band does not vanish,
in contrast to the case of the direct Auger mechanism.  Since the matrix
element for normal processes (${\bf G}=0$) does not vanish for $K=0$, the
contribution of Umklapp processes should be relatively small, and we neglect
them.  In this case the angular momentum of the colliding excitons does not
put any restriction on Auger collisions.  Following the same procedure as
before, we calculate all Feynman diagrams (Fig. 3).  We assume for simplicity
that the lattice temperature is very low, and only spontaneous phonon emission
is possible.

      For the processes C, C$^\prime$ and D, D$^\prime$ of Fig. 3 with two
excitons of momentum ${\bf K}$ and ${\bf P}$ colliding, giving an
electron of momentum ${\bf k_e}$, a hole of momentum ${\bf k_h}$ and a
longitudinal optical (LO) phonon of momentum ${\bf Q}$ and energy
$\hbar \omega_{{\bf Q}}$, denoted by the dashed line, the matrix element is
\begin{eqnarray}
     M_C+M_{C'}+M_D+M_{D'} =
   V_{\text{eff}}({\bf K} - {\bf Q}) \,
     (\phi_{{\bf k_h} - {\bf P}/2} - \phi_{{\bf k_e} - {\bf P}/2})
\phantom{XXXXXXXXXXXXXXXXX}
\nonumber\\
\phantom{XXXXXX}
     \times  \sum_{{\bf q}, n}
    \phi_{{\bf q}} \left(
   \frac {\langle u_{v,{\bf q} - {\bf K}/2} |
  u_{n,{\bf q} + {\bf K}/2 - {\bf Q}} \rangle
     \langle u_{n,{\bf q} + {\bf K}/2 - {\bf Q}} | H_{\text{LO}} |
u_{c,{\bf q} + {\bf K}/2} \rangle } {\varepsilon_{c,{\bf q} + {\bf K}/2} -
    \varepsilon_{n,{\bf q} + {\bf K}/2 - {\bf Q}} - \hbar \omega_{{\bf Q}}}
\right.
\phantom{XXXXXXXXXXXXXX}
\nonumber \\
 + \left.
  \frac { \langle u_{v,{\bf q} - {\bf K}/2} | H_{\text{LO}} |
	u_{n,{\bf q} - {\bf K}/2 + {\bf Q}} \rangle
    \langle u_{n,{\bf q} - {\bf K}/2 + {\bf Q}} |
	u_{c, {\bf q} + {\bf K}/2} \rangle}
   {\varepsilon_{v,{\bf q} - {\bf K}/2} -
 \varepsilon_{n,{\bf q} -{\bf K}/2 + {\bf Q}}+\hbar \omega_{{\bf Q}}} \right)
      \delta_{{\bf k_e} + {\bf k_h} + {\bf Q}, {\bf K} + {\bf P}},
\phantom{XXXXX}
\label{61}
\end{eqnarray}
where $H_{\text{LO}}$ is the carrier (i.e., electron or hole) -- LO-phonon
interaction. The first term in the sum refers to processes C and
C$^\prime$, and the second to processes D and D$^\prime$. The sum over $n$
includes the two odd-parity bands of Cu$_2$O.
We again use ${\bf k} \cdot {\bf p}$ perturbation theory to calculate the
overlap integrals in the above matrix element.  For example,
\begin{eqnarray}
 \langle u_{v,{\bf q} - {\bf K}/2}| u_{n, {\bf q} + {\bf K}/2 - {\bf Q}}\rangle
  \approx \frac \hbar m \frac {\langle u_{v, {\bf q} - {\bf K}/2}| ({\bf Q} -
{\bf K}) \cdot {\bf p} | u_{n, {\bf q} - {\bf K}/2} \rangle } {\varepsilon_{n,
 {\bf q} - {\bf K}/2} - \varepsilon_{v, {\bf q} - {\bf K}/2} }.
\label{63}
\end{eqnarray}

      The two terms in the sum of Eq.  (\ref{61}) are closely related to the
matrix elements for radiative phonon-assisted exciton recombination.  We show
the radiative recombination processes of an exciton in Fig. 3, diagrams F and
F$^\prime$, where an exciton recombines emiting a phonon of momentum ${\bf Q}$
and a photon of momentum ${\bf k_{\gamma}}$.  The similarity of the upper part
of diagrams C, C$^\prime$, D, and D$^\prime$ with F, and F$^\prime$, enables
us to extract information about the phonon-assisted Auger mechanism.
Experimentally, the primary luminescence mechanism of ortho excitons in
Cu$_2$O is via phonon-assisted recombination involving the $\Gamma_{12}^{-}$
optical phonon.  Other phonon-assisted recombination processes of the ortho
excitons are at least 30 times weaker.  For the para excitons the only allowed
optical phonon recombination mechanism is via a $\Gamma_{25}^{-}$ phonon, and
the integrated intensity is about 500 times weaker than the $\Gamma_{12}^{-}$
ortho-exciton phonon-assisted mechanism.  This difference is due to the band
structure of the material and the symmetry of the bands that are involved in
the transition.  Group theory allows both odd-parity bands to assist the
ortho-exciton radiative recombination, so for the orthoexcitons both processes
F and F$^\prime$ are allowed.  In para-exciton recombination, in contrast,
only the deep valence band can be the intermediate state (diagram F$^\prime$),
as illustrated in Fig. 6 \cite{Birm}.  This analysis is valid only for
non-stressed crystals; if stress is applied the mixing between the
$\Gamma_7^+$ and the $\Gamma_8^+$ valence bands, dependent on the orientation
of the applied stress, is expected to affect the phonon-assisted radiative
recombination processes.

    In the Auger problem this symmetry argument implies that for ortho-exciton
recombination, the rate due to processes C and C$^\prime$ is much larger than
the rate due to D and D$^\prime$.  For para excitons, processes C and
C$^\prime$ are forbidden and D and D$^\prime$ contribute negligibly to the
decay rate.  To estimate the ratio of the magnitudes of the matrix elements
for the faster ortho-exciton ($M_C+M_{C'}$) and the para-exciton
($M_D+M_{D'}$) Auger phonon-assisted mechanisms due to this effect we note
that the only difference in this case as compared to the radiative
recombination comes from the energy denominator in Eq.  (\ref{63}).  An
estimate of this ratio is
\begin{eqnarray}
      \left| \frac {M_C+M_{C'}} {M_D+M_{D'}} \right|  \approx
    \sqrt \frac {\Gamma_{o,\text{rad}}} {\Gamma_{p,\text{rad}}}
\,   \frac {\varepsilon_{c',0} - \varepsilon_{v,0}}
   {\varepsilon_{c,0} - \varepsilon_{v',0}}
     \approx \sqrt {500}
\,   \frac {\varepsilon_{c',0} - \varepsilon_{v,0}}
  {\varepsilon_{c,0} - \varepsilon_{v',0}},
\label{relat}
\end{eqnarray}
where $\Gamma_{o,\text{rad}}$ is the total decay rate of ortho excitons
due to the $\Gamma_{12}^{-}$ phonon and $\Gamma_{p,\text{rad}}$ is the
total decay rate of para excitons due to the $\Gamma_{25}^{-}$ phonon.
As can be seen from the energy differences
in the denominator of Eq. (\ref{relat}) using the energy levels of Cu$_2$O,
$|(M_C+M_{C'})/(M_D+M_{D'})| \approx 3.0 \times\sqrt{500}$.
The only assumption we have made in writing Eq. (\ref{relat}) is the
reasonable one that
the oscillator strengths between the bands $c'$, $v$, and $c$, $v'$ are
comparable.
Thus, the relative decay rate due to processes C, C$^\prime$ as compared
with the decay rate of D, D$^\prime$ is $\approx 5 \times 10^3$. Since,
as we show below, all other diagrams (of the form of E in Fig. 3) for
the phonon-assisted Auger mechanism can be neglected, we conclude that
para excitons have a negligible phonon-assisted Auger decay rate (in para-para
collisions), compared with the ortho excitons, in agreement with the observed
long para-exciton lifetime when no ortho excitons are present. Since the
$\Gamma_{12}^{-}$ optical phonon dominates the phonon-assisted Auger decay
process of ortho excitons, we neglect the contribution of all other phonons.

      Expanding the overlap integral $\langle u_{v,{\bf q}-{\bf K}/2}|
u_{c',{\bf q}+{\bf K}/2-{\bf Q}}\rangle$ in powers of ${\bf K} - {\bf Q}$,
using ${\bf k} \cdot {\bf p}$ perturbation theory,
we see that the matrix element for processes C and C$^\prime$ goes as
\begin{eqnarray}
    M_C+ M_{C'} \sim
   \frac 1 {|{\bf K} - {\bf Q}|^2}
  (\phi_{{\bf k_e} - {\bf P}/2 - {\bf K}  + {\bf Q}} -
 \phi_{{\bf k_e} - {\bf P}/2})
 ({\bf K} - {\bf Q}) \cdot {\bf p}_{v,c'},
\label{64}
\end{eqnarray}
so the matrix element $M_C+M_{C'}$ can be treated as constant. We estimate
it using Eqs. (\ref{61}) and (\ref{63}) as
\begin{eqnarray}
     M_C+M_{C'} \approx \frac 1 {\Omega} ({\bf K} - {\bf Q}) \cdot
  \left( \frac {\partial \phi_{{\bf p}}} {\partial {\bf p}} \right)_{{\bf k_e}}
 \frac {4 \pi e^2} {\epsilon_{\infty} |{\bf K} - {\bf Q}|^2} \,
    \left( \sum_{{\bf q}} \phi_{{\bf q}} \right)
   \frac \hbar m \frac {{\bf p}_{v,c'} \cdot
 ({\bf K} - {\bf Q})} {(\varepsilon_{c',0} - \varepsilon_{v,0})}
 \frac {\langle u_{c',0} | H_{\Gamma_{12}^{-}} | u_{c,0} \rangle}
     {(\varepsilon_{c,0} - \varepsilon_{c',0})}
   \, \delta_{{\bf k_e} + {\bf k_h} + {\bf Q}, {\bf K} + {\bf P}}.
\label{65}
\end{eqnarray}
The matrix element $\langle u_{c',0} | H_{\Gamma_{12}^{-}} | u_{c,0}
\rangle$ equals $D_{\Gamma_{12}^{-}} [ \hbar/(2 \rho
\Omega \omega_{\Gamma_{12}^{-}}) ]^{1/2}$ in magnitude, where
$\omega_{\Gamma_{12}^{-}}$ is the zone center frequency of the
specific phonon, $\approx 13.8$ meV, $\rho$ is the mass density of
the material, and $D_{\Gamma_{12}^{-}}$
is the deformation potential involving the $\Gamma_{12}^{-}$ optical
phonon-mediated interband transition between the $\Gamma_6^{+}$ and the
$c'$ band. Therefore, Eq. (\ref{65}) can be written as
\begin{eqnarray}
      M_C+M_{C'} \approx 128 \pi \frac {e^2 a_B} {\epsilon_{\infty} {\Omega}}
\,      \frac \hbar m
   \frac {|{\bf p}_{v,c'}|} {(\varepsilon_{c',0} - \varepsilon_{v,0})}
      \frac {k_e a_B} {[1+(k_e a_B)^2]^3}
   \frac {D_{\Gamma_{12}^{-}}} {(\varepsilon_{c,0} - \varepsilon_{c',0})}
\left( \frac {\hbar} {2 \rho \Omega \omega_{\Gamma_{12}^{-}}} \right)^{1/2}
\,  \delta_{{\bf k_e} +{\bf k_h} + {\bf Q}, {\bf K} + {\bf P}}.
\label{66}
\end{eqnarray}
The deformation potential $D_{\Gamma_{12}^{-}}$ has not been measured directly,
but we can extract an estimate from Ref. \cite{Mys}. We use the
measured para-exciton lifetime of 13 $\mu$s at a temperature of 10 K and the
relative integrated intensities of the $\Gamma_{12}^{-}$
ortho-exciton phonon-assisted mechanism with respect to the $\Gamma_{25}^{-}$
para-exciton phonon-assisted mechanism (approximately 500, as
mentioned earlier), to fit the decay rate of the ortho excitons due
to the phonon-assisted recombination, 26 ns at the same temperature of 10 K.
This analysis implies that $D_{\Gamma_{12}^{-}} \approx 2.5$ eV/\AA.

     For the processes of the form of E shown in Fig. 3,
which correspond to the diagrams B and B$^\prime$ (direct Auger process)
with additional emission of a $\Gamma_{12}^{-}$ phonon, we have
\begin{eqnarray}
     M_E= \sum_{{\bf q}} \phi_{{\bf q}} \phi_{{\bf k_h} - {\bf P}/2}
   V_{\text{eff}}({\bf k_e} - {\bf q} - {\bf K}/2)
\frac { \langle u_{v,{\bf q} - {\bf K}/2}|u_{c',{\bf k_e} - {\bf K}} \rangle
	\langle u_{c',{\bf k_e} - {\bf K}} |
    H_{\Gamma_{12}^{-}} | u_{c,{\bf P} - {\bf k_h}}
\rangle}
{ \varepsilon_{c,{\bf k_h} - {\bf P}} - \varepsilon_{c',{\bf k_e} - {\bf K}}
	       - \hbar \omega_{{\bf Q}} } \,
\delta_{{\bf k_e} + {\bf k_h} + {\bf Q} , {\bf K} + {\bf P}}.
\label{E}
\end{eqnarray}
Since the overlap integral is less than unity, an upper bound on $M_E$ is
\begin{eqnarray}
M_E \le
    \left| \phi_{{\bf k_h}}
   \frac {\langle u_{c',0} | H_{\Gamma_{12}^{-}} | u_{c,0} \rangle}
 { \varepsilon_{c,0} - \varepsilon_{c',0} }
  \sum_{{\bf q}} \phi_{{\bf q}}
   V_{\text{eff}}({\bf k_e} - {\bf q}) \,
 \delta_{{\bf k_e} + {\bf k_h} + {\bf Q}, {\bf K} + {\bf P}} \right|.
\label{Eest}
\end{eqnarray}
The sum over ${\bf q}$ in Eq. (\ref{Eest}), the convolution of
the wave function of the relative electron-hole motion $\phi_{\text{rel}}(x)$
times the Coulomb interaction $V_{\text{eff}}(x)$, can be calculated
analytically, yielding
\begin{eqnarray}
   M_E \le
 \frac 1 {\Omega^{1/2}} \phi_{{\bf k_h}} \frac {D_{\Gamma_{12}^{-}}}
     {(\varepsilon_{c,0} - \varepsilon_{c',0})}
 \left( \frac {\hbar} {2 \rho \Omega \omega_{\Gamma_{12}^{-}}} \right)^{1/2}
    \frac {4 {(\pi a_B)}^{1/2} e^2} {\epsilon_{\infty} [1+(k_e a_B)^2]}
\,      \delta_{{\bf k_e} + {\bf k_h} + {\bf Q}, {\bf K} + {\bf P}}.
\label{En}
\end{eqnarray}

    Let us proceed to the calculation of the decay rate. If $M_{\text{ph}}$
is the matrix element for the phonon-assisted process, the decay rate of two
colliding excitons with momenta ${\bf K}$ and ${\bf P}$ giving an electron of
momentum ${\bf k_e}$, a hole of momentum ${\bf k_h}$ with simultaneous
emission of an optical phonon of momentum ${\bf Q}$, is
\begin{eqnarray}
     \Gamma_{{\bf K},{\bf P}}=
\frac {2 \pi} \hbar \sum_{{\bf k_e}, {\bf k_h}, {\bf Q}}
	|M_{\text{ph}}|^2 (1- n_{c,{\bf k_e}})
      (1 - n_{v,{\bf k_h}}) \delta (E_{{\bf K}} + E_{{\bf P}} -
  \hbar \omega_{{\bf Q}} - \varepsilon_{c,{\bf k_e}} -
    \varepsilon_{v,{\bf k_h}}).
\label{67}
\end{eqnarray}
We neglect the inverse process at very low lattice temperature, where
no phonons are present.

     The contribution of the (dominant) processes C and C$^\prime$ to the
Auger phonon-assisted decay rate, defined in Eq. (\ref{53}),
is, using Eq. (\ref{66}),
\begin{eqnarray}
 \tau_{A,\text{ph}}^{-1} \approx \frac {2^{16} \times 10^{-2}} {\pi}
  \frac {\mu e^4} {2 \hbar^3 \epsilon_{\infty}^2}
    \frac {\hbar^4} {m \rho a_B^7} \, \, n a_B^3
       \frac {|{\bf p}_{c',v}|^2} m
	  \frac {(D_{\Gamma_{12}^{-}} a_B)^2}
	    {(\varepsilon_{c,0} - \varepsilon_{c',0})^2
   (\varepsilon_{c',0} - \varepsilon_{v,0})^2 \hbar \omega_{\Gamma_{12}^{-}}}.
\label{69}
\end{eqnarray}
In the approximation of Eq. (\ref{64}), where the matrix element
is constant, $\tau_{A,\text{ph}}^{-1}$ is temperature
independent at low lattice temperature and the corresponding inverse scattering
time is proportional to the density of the total number of excitons,
$\tau_{A,\text{ph}}^{-1} \sim n $. If we measure the density in units of
10$^{18}$ cm$^{-3}$, Eq. (\ref{69}) gives the decay rate
\begin{eqnarray}
 \tau_{A,\text{ph}}^{-1} \approx 0.5 \, n \, \text{ns}^{-1},
\label{611}
\end{eqnarray}
which is the order of magnitude of the observed decay rate,
but not numerically accurate.

      For process E, the approximate expression on the right side of
(\ref{En}) for the matrix element yields a lower bound on the contribution
$\tau_{A,\text{ph},E}$ to the Auger phonon-assisted scattering time,
\begin{eqnarray}
	\tau_{A,\text{ph},E}^{-1} \le
	\frac {2^{10} \times 10^{-4}} \pi
\frac {\mu e^4} {2 \hbar^3 \epsilon_{\infty}^2}
	     \frac {\hbar^2} {\rho a_B^5} \, \, n a_B^3
	   \frac {(D_{\Gamma_{12}^{-}} a_B)^2}
 {(\varepsilon_{c,0} - \varepsilon_{c',0})^2 \hbar \omega_{\Gamma_{12}^{-}}}.
\label{Etn}
\end{eqnarray}
Numerically $\tau_{A,\text{ph},E}^{-1} \le 9 \times 10^{-3} \, n$
ns$^{-1}$, with $n$ measured in units of $10^{18}$ cm$^{-3}$, which
is comparable to the decay rate of the para excitons,
i.e., the smallest decay rate of our problem. Processes of the
form of E are in fact much slower than this rate, since we have assumed
in (\ref{Etn}) that the overlap integral is unity. Furthermore,
the process E$^\prime$, of the form of E with the Coulomb exchange
coupling instead to the positively charged hole, enters with a change of sign,
partially cancelling E.

\section{Effects of identity on the Auger decay rate}

      We turn now to the problem of the identity between the
electrons and holes of the two excitons. The total wave function of
two excitons of momenta ${\bf K}$ and ${\bf P}$ with angular momentum
functions $\chi_{\alpha}$ and $\chi_{\beta}$ respectively, is
\begin{eqnarray}
   \Psi_{i,\text{tot}} =\frac 1 2 [ \Psi_{{\bf K}}(1,2)
\chi_{\alpha}(1,2) \Psi_{{\bf P}}(1',2') \chi_{\beta}(1',2')
      &-& \Psi_{{\bf K}}(1',2) \chi_{\alpha}(1',2)\Psi_{{\bf P}}(1,2')
    \chi_{\beta}(1,2')
\nonumber \\
     -\Psi_{{\bf K}}(1,2')
    \chi_{\alpha}(1,2')\Psi_{{\bf P}}(1',2)  \chi_{\beta}(1',2)
&+& \Psi_{{\bf K}}(1',2') \chi_{\alpha}(1',2')\Psi_{{\bf P}}(1,2)
    \chi_{\beta}(1,2) ] ,
\label{56}
\end{eqnarray}
where the variables $1,1',3$ refer to electrons and $2,2',4$ to holes.
The wave fuction of the final state is
\begin{eqnarray}
      \Psi_{f,\text{tot}}= \Psi_{{\bf k_e},{\bf k_h}}(3,4) \chi_{\gamma}(3,4).
\label{57}
\end{eqnarray}
The calculation we have presented has been done without effects of identity
taken into account. Thus, if the initial state in the matrix elements
$M_A, M_{A'}, M_C$ and $M_{C'}$ is the first term of Eq. (\ref{56}), the
second term of Eq. (\ref{56}) gives the initial state in the matrix
elements $M_B, M_{B'}, M_D$ and $M_{D'}$; the last two give exchange terms.
Since the phonon-assisted processes C and C$^\prime$ dominate
the Auger collisions, we examine here the effect of identity of the
electrons and the holes on the corresponding matrix elements $M_C$ and
$M_{C'}$.

    The Coulomb interaction does not produce any spin-flips, so the angular
momentum functions $\chi_i$ factor out in all the matrix elements, independent
of the spatial part of the calculation.  For ortho-para and for ortho-ortho
collisions with different $J_z$, only the first or the last term of Eq.
(\ref{56}) contributes to the matrix element, because of the orthogonality of
the angular momentum wave functions.  The actual matrix element, therefore,
for the case that the effect of identity has been taken into account is 1/2
(from the normalization constant) times the result of our calculation from the
previous section.  For ortho-ortho collisions with the same $J_z$, both the
first as well as the last term in Eq.  (\ref{56}) contribute to the matrix
element and therefore the numerical factor which multiplies the matrix element
in this case is 1/2+1/2=1.

     In a gas of orthoexcitons with random $J_z$, there are nine
possible combinations, with respect to their angular momentum, that two
ortho excitons collide. Statistically 3/9 of the collision events are between
ortho excitons with the same $J_z$, and 6/9 between ortho excitons with
different $J_z$. Consequently, if $C_A$ ($\approx 0.5$ from Eq. (\ref{611}))
is the proportionality factor in our calculation without taking into
account the effect of identity, the effect of identity on the decay rate
is to multiply the original rate by a factor $(3/9) \, 1 + (6/9) \,
(1/2)^2=1/2$ for ortho-ortho collisions and $(1/2)^2$ for ortho-para
collisions. Thus for the dominant phonon-assisted Auger process
\begin{eqnarray}
  - \frac 1 {N_o} \left( \frac {\partial N_o}
      {\partial t} \right)_{A,\text{ph}}^o &\equiv&
  \frac 1 {\tau_{A,\text{ph}}^o}
  = \frac {C_A} 2 \left( n_o + \frac 1 2 n_p \right),
\label{58}
\\
    - \frac 1 {N_p} \left( \frac {\partial N_p}
   {\partial t} \right)_{A,\text{ph}}^p &\equiv&
  \frac 1 {\tau_{A,\text{ph}}^p} = \frac {C_A} 4 n_o,
\label{59}
\end{eqnarray}
where $N_o$ and $N_p$ are the numbers of ortho and para excitons, respectively.
To determine the {\it net\/} loss of ortho and para excitons due to the Auger
process, we take into account the fact that the Auger-ionized excitons
reform in random angular momentum states, and find
\begin{eqnarray}
- \frac 1 {N_o} \left( \frac {\partial N_o}
      {\partial t} \right)_{A,\text{net}}^o
  &=& \frac {C_A} {16} \left(5 n_o + {n_p} \right),
\label{6760}
  \\
      - \frac 1 {N_p} \left( \frac {\partial N_p}
    {\partial t} \right)_{A,\text{net}}^p
  &=& \frac {C_A} {16} n_o \left( 3 - \frac{n_o} {n_p} \right).
\label{6761}
\end{eqnarray}

\section{Experimental observations -- Conclusions}

    The two basic experimental observations related to the Auger process are
shown in Figs. 2 and 7. Figure 2 shows the total decay rate per particle of
the ortho excitons as function of the ortho-exciton density.  As seen from
this graph, the total decay rate is approximately proportional (the exact
power is 0.8) to the ortho-exciton density.  In this experiment a mode-locked,
cavity dumped Argon-ion laser produces nanojoule pulses with about a 100-ps
length.  For a photoluminescence time resolution of about 100 to 300 ps, this
excitation pulse is effectively a delta-function in time, and the evolution of
the system is observed without creation by the laser of further particles.
The excitons form on a timescale shorter than the detection limits, within a
few nanoseconds.  The much weaker radiative efficiency of the para excitons
makes it difficult to observe them during this short time period.  As
mentioned earlier, this experiment shows quantum saturation of the ortho
excitons, i.e., they move along an adiabat ($- \mu_o/ k_BT \approx 0.2$, where
$\mu_o$ is the chemical potential of the ortho excitons) very close to the
condensation line, without condensing; along adiabats $n_o \sim T^{3/2}$.  Our
result, Eq.  (\ref{6760}), gives an approximate linear dependence of the Auger
decay rate per ortho exciton as function of the ortho-exciton density.
Experimentally the Auger decay rate cannot be measured directly, since
ortho-para interconversion \cite{op} contributes almost as much as the Auger
to the total rate shown in Fig. 2. Since the decay rate due to the conversion
of ortho excitons to para excitons is approximately $\sim T^{3/2}$, which is
$\sim n_o$ along an adiabat, the contribution of the Auger decay rate to the
total rate experimentally is also $\sim n_o$, in accordance with our
theoretical calculation.  In the numerical simulation in Ref.  \cite{kbw}, the
value of 0.40 ns$^{-1}$ for the Auger decay constant $C_A$ (with the density
measured in units of 10$^{18}$ cm$^{-3}$) reproduces the experimental results;
this value should be compared with 0.5 ns$^{-1}$ from Eq.  (\ref{611}).

    If the mode-locker is removed from the laser, the cavity-dumped mode
provides 10-ns long pulses with about an order of magnitude more energy.
Figure 7(a) shows data from lightly stressed crystals with such a long-pulse
excitation \cite{cond}.  The laser profile (triangles), the number of ortho
excitons in the lowest ortho-exciton level (open circles), and the number of
para excitons (black dots) are shown as function of time in this figure.  The
stress splits the triply degenerate ortho-exciton level into three components
and only the lowest of the three is significantly populated, leading to a
closer proximity to the condensation line than in the unstressed case.  No
ortho-exciton condensation is observed.  Under these conditions the
para-exciton density was determined from the relative intensities of ortho and
para excitons, combined with the spectroscopically determined density of the
ortho excitons.  This analysis yields the striking evidence that the para
excitons condense shown in Fig.~7(b), a graph of the corresponding
trajectories for the ortho excitons (open circles) and the para excitons
(black dots) in the density-temperature plane.  The straight line here is the
condensation phase boundary, which is identical for para excitons and ortho
excitons in the stressed case.  The para excitons in this case are Bose
condensed for times later than 8 ns.  A crucial feature of the para excitons
in these data which is related with the Auger process is their significantly
smaller decay rate than that of the ortho excitons.  More specifically, their
lifetime is on the order of 100 ns at a density $10^{19}$ cm$^{-3}$, while the
lifetime of the noncondensed ortho excitons ($\approx$ 0.1 nsec) is consistent
with the phonon-assisted Auger and ortho-para conversion mechanisms discussed
earlier, in the case of short-pulse excitation.  The para-exciton decay time
of 100 ns refers to late times, when the production of excitons due to the
laser is negligible, the ortho-exciton density is very low, and essentially
only para excitons exist.  This observation shows that para-para Auger
collisions lead to a very low decay rate.  As we showed in Sec.  IV, the
reason for this low rate is that the band with the required symmetry which
assists the para-para phonon-assisted Auger processes is a deep valence band.
See Fig. 6. In contrast, for ortho-ortho, or ortho-para collisions the
relevant band is very close to the conduction band.  In the numerical
simulation in Ref.  \cite{kbw} of the time dependence of excitons, we have
shown that the ortho excitons move along the phase boundary without crossing
it while the para excitons condense, as a result of the Auger heating of the
othro excitons and phonon cooling of both the ortho and para excitons.

    The result of high uniaxial stress $\approx 3.6$ kbar on the Auger process
has been resolved in Ref.  \cite{stress}.  For comparison, in the data of Fig.
7 the stress is $\approx$ 0.36 kbar; in the data of Fig. 2 there is no applied
stress.  The experiment in Ref.  \cite{stress} shows clearly that para-para
Auger collisions are allowed in stressed crystals.  In this experiment laser
light produces the excitons away from the well, which drift a distance of
$\approx$ 400 $\mu$m in the stress gradient to the shear-stress maximum where
they are confined.  The number of ortho excitons in the well should be
negligible, since the relatively short lifetime of ortho excitons precludes
travel over $400 \, \mu$m distances; also the expected number of ortho
excitons due to thermal excitations is negligibly small under the specific
experimental conditions ($T \approx$ 2 K).  Experimentally, however, the
measured number of ortho excitons is approximately proportional to the square
of the para-exciton number, and the maximum para-exciton density goes as the
square root of the laser power.  These observations have been attributed to
the Auger decay of para excitons \cite{stress}.  In this case that the crystal
is highly-stressed, the symmetry selection rules we have used in both the
phonon-assisted, as well as the direct processes are expected to break down.
An example of this symmetry breaking is that the (zero stress-forbidden)
direct recombination line of para excitons has been observed clearly in
stressed crystals; as far as we know, however, the result of stress on
phonon-assisted recombination has not been investigated yet.

    Stress strongly influences the Auger process, starting with the
phonon-assisted mechanism; both the dominant phonon-assisted Auger process
involving the $\Gamma_{12}^{-}$ phonon as well as the direct Auger become
allowed for para-para collisions in stressed crystals, for specific directions
of the stress.  More specifically, group theory predicts \cite{pstr} that if
the uniaxial stress is along the $C_2$ or the $C_3$ crystallographic axes, the
mixing of the $\Gamma_{7}^{+}$ with the $\Gamma_{8}^{+}$ band allows the
$\Gamma_{12}^{-}$ LO-phonon to assist the process.  In the specific experiment
of Ref.  \cite{stress} the applied stress is primarily along the $C_4$ axis,
but it also has a component in the $C_2$ direction.  Moreover, uniaxial stress
makes the direct Auger process (dominated by the processes shown in diagrams A
and A$^{\prime}$) weakly allowed.  Both processes are responsible for the
presence of ortho excitons in the quantum well, which are generated by the
reformation of Auger-ionized para excitons in random angular momentum states.
In both cases the orientation of the uniaxial stress is very important, since
it determines the selection rules in the deformed crystal \cite{pstr}.

    In summary, we have calculated the direct and phonon-assisted Auger decay
rate, incorporating the band structure of Cu$_2$O, which plays an essential
role because of the following effects:  1) the same (positive) parity of the
two bands, 2) the fact that the valence band is not a pure spin state and 3)
the symmetry and the location of the negative-parity bands with respect to the
conduction and valence bands which form the yellow excitons.  The inverse
Auger decay times for the direct as well as the much faster phonon-assisted
Auger mechanism are proportional to the exciton density.  Finally, we have
shown that the para-exciton recombination vertex is either forbidden (direct
processes), or negligible (phonon-assisted processes), and consequently
para-para Auger collisions have a negligible Auger decay rate in non-stressed
crystals.

\acknowledgments
     This work was supported by NSF Grants Nos. DMR91-22385 and PHY94-21309.
     Helpful comments from Y.C. Chang,
     K. O'Hara, L. O'Suilleabhain, D.W. Snoke and J.P. Wolfe are gratefully
     acknowledged. G.M.K. wishes to thank the Research Center of Crete, Greece
     for its hospitality.

\figure{FIG. 1.
     Two classes of diagrams for the direct Auger non-radiative
recombination. Time progresses from left to right. The initial state
contains two excitons of momenta ${\bf K}$ and ${\bf P}$, and the final state
contains an ionized electron and hole with momenta ${\bf k_e}$ and ${\bf k_h}$,
respectively. The dashed line denotes the Coulomb interaction. Processes
A and A$^{\prime}$ are the dominant ones and they are nonzero only if the
recombined particle is an ortho exciton.}

\figure{FIG. 2.
      The {\it total\/} decay rate \cite{augn} of ortho excitons versus the
ortho-exciton density in unstressed crystals. Figure reproduced with the
kind permission of J.P. Wolfe.}

\figure{FIG. 3.
The phonon-assisted Auger process (diagrams C, C$^\prime$, D, D$^\prime$
and E), and the phonon-assisted radiative recombination process (F and
F$^\prime$).  The phonon is denoted by a dashed line.  Processes D and
D$^{\prime}$ are the dominant ones because of the band structure of Cu$_2$O.
In these, the recombined exciton can only be an ortho exciton.  Similarly,
diagram F is dominant compared to F$^\prime$; the recombined exciton can only
be an ortho exciton.}

\figure{FIG. 4.
     Virtual annihilation diagram of an exciton, possible only for
pure spin-singlet excitons.}

\figure{FIG. 5.
 Schematic band structure of Cu$_2$O showing the conduction $\Gamma_6^{+}$ band
and the $\Gamma_7^{+}$, $\Gamma_8^{+}$ valence bands (split by the spin-orbit
splitting) which form the yellow and green exciton series, respectively.}

\figure{FIG. 6.
     Phonon-assisted radiative recombination processes. The ortho excitons
are allowed to participate in both, but for the para excitons only the one
on the right side is allowed. The bands $c,v$ are the $\Gamma_6^{+}$ and the
$\Gamma_7^{+}$ bands which form the yellow excitons and are
separated by 2.17 eV. The bands $c'$ and $v'$ are the only odd-parity bands in
Cu$_2$O. The energy difference between $c'$ and $c$ is $\approx$ 449 meV and
between $v$ and $v'$ is $\approx$ 5.6 eV. The proximity of band $c'$ to
band $c$ makes both the radiative recombination \cite{Birm} and the
phonon-assisted Auger decay processes much faster for ortho excitons
than para excitons.}

\figure{FIG. 7.
     (a) Data from lightly stressed crystals with long-pulse
(10-ns) excitation \cite{cond}.  The laser profile (triangles), the number of
ortho excitons in the lowest ortho-exciton level (open circles), and the number
of para excitons (black dots) as function of time.  The para excitons show a
significantly smaller decay rate.  (b) Corresponding trajectories for
ortho excitons (open circles) and para excitons (black dots) in the
density-temperature plane.  The straight line is the condensation phase
boundary, which is identical for para excitons and ortho excitons in the
stressed case.  Note that the para excitons are in the condensed region at
times later than 8 ns. Figure reproduced with the kind permission of J.P.
Wolfe.}

\end{document}